\documentclass[aps,showpacs]{revtex4}

\usepackage{graphicx}
\usepackage{dcolumn}
\usepackage{bm}
\usepackage{amsmath}
\newcommand{\be}{\begin{equation}}
\newcommand{\ee}{\end{equation}}
\newcommand{\bea}{\begin{eqnarray}}
\newcommand{\eea}{\end{eqnarray}}
\newcommand{\beb}{\begin{eqnarray*}}
\newcommand{\eeb}{\end{eqnarray*}}
\begin{document}
\preprint{LPTMS-2007-XXX}

\title{Fermions out of Dipolar Bosons in the lowest Landau level}

\author{B.~Chung and Th.~Jolic\oe ur}

\affiliation{Laboratoire de Physique Th\'eorique et Mod\`eles
Statistiques, Universit\'e Paris-Sud, Orsay, France}

\date{December 19, 2007}
\begin{abstract}
In the limit of very fast rotation atomic Bose-Einstein condensates
may reside entirely in the  lowest two-dimensional Landau level (LLL). For small enough
filling factor of the LLL, one may have formation
of fractional quantum Hall states. We investigate the case of bosons with
dipolar interactions as may be realized with Chromium-52 atoms. We show
that at filling factor equal to unity the ground state is a  Moore-Read
(a.k.a Pfaffian) paired state as is the
case of bosons with purely s-wave scattering interactions. This Pfaffian state
is destabilized when the interaction in the s-wave channel is small enough and
the ground state is a stripe phase with unidimensional density modulation.
For filling factor 1/3, we show that there is formation of a Fermi sea
of ``composite fermions''. These composites are made of one boson bound with three vortices.
This phase has a wide range of stability and the effective mass of the fermions
depends essentially only of the scattering amplitude in momentum channels larger or
equal to 2. The formation of such a Fermi sea opens up a new possible  route
to detection of the quantum Hall correlations.
\end{abstract}

\maketitle
\section{Introduction}
The quantum behavior of a Bose-Einstein condensate is readily apparent
in the response to rotation. Instead of solid-body rotation as in a viscous fluid,
there is formation of lines of quantized vorticity. These vortices
then form a regular lattice eventually distorted by the container, i.e.
the trapping potential. In the case of structureless bosons interacting by purely s-wave
scattering, the equilibrium vortex lattice is the celebrated triangular Abrikosov
lattice~\cite{Madison00,Aboshaeer01}. Many experiments have been able to observe and study in detail
this vortex lattice. Its structure and dynamics have been a very active
field with fruitful interplay between theory and experiments\cite{BlochRMP}.
The angular momentum of the gas in this regime is of the order of
the number of particles times the number of vortices. From theoretical grounds,
it is expected that a physically different regime is reached when
the angular momentum is raised up to the order of the square of the number of particles.
In this case, various quantum phases at zero temperature
 may appear~\cite{Wilkin98,Wilkin00,Cooper01}, 
related to the 
fractional quantum Hall liquids already known from the physics of two-dimensional electron systems.
These phases exist when the gas enters a bidimensional regime, either 
forced by an external confining potential along the rotation axis or induced
by the centrifugal force. The mere existence of these phases is due to the
formation of Landau levels in which kinetic energy is quenched and
all the physics is determined by the interaction potential between atoms.
When the centrifugal force compensates the harmonic trapping potential, one is left with
the Coriolis force which is formally equal to a fictitious magnetic field
acting upon the neutral atoms. In a two-dimensional situation, the fictitious field
leads to one-body energy levels that are degenerate, the Landau levels (LLs).
The systematics of the apparition of the phases is ruled by the 
so-called filling factor $\nu$, i.e. the number of atoms divided by the number
of states in the lowest Landau level (LLL). The quantum Hall phases
are expected at filling factors of order unity. Even when all bosons
reside entirely in the LLL, there will be a vortex lattice when the
filling factor is large enough~\cite{schweikhard04}. This lattice may 
melt~\cite{Cooper01}
 and be replaced
by quantum Hall liquids in the region $\nu\approx 1$. The case of
bosons with pure s-wave scattering is special. Indeed one may
model the interaction in appropriate circumstances by a pure delta
function interaction whose coefficient is proportional to the s-wave scattering length.
Right at $\nu =1/2$ it is the celebrated
Laughlin wavefunction which is the exact ground state of the system of interacting bosons.
In this case for filling factors smaller that $\nu =1/2$ interactions play
absolutely no role in the ground state. Indeed the ground state becomes increasingly degenerate
with increasing angular momentum, being given by the product of an arbitrary symmetric polynomial
times the Laughlin wavefunction. In fact the gas is noninteracting in this regime since
all pairs of atoms have a relative angular momentum which is at least two.

It has been recently possible to create condensates of chromium atoms which have dipolar
interactions~\cite{Griesmaier05,Stuhler05,Lahaye07,Menotti07,Menotti07-1,Menotti07-2}. 
This opens up the possibility to study the change of interaction potential
in many-body problems. The problematic of response to rotation can be again investigated
with this new interaction potential. For example, the vortex lattice has been shown
to display a much richer phase structure~\cite{Komineas06,Cooper05,Zhang05}. 
Beyond triangular and square arrangements
of vortices, there is also evidence~\cite{Cooper05,Rezayi05,Cooper07}
 of phases with unidimensional density modulation pattern
and regular two-dimensional arrays of bubbles containing more than one atom. The physics
of atoms in the LLL is also affected by the dipolar interactions. 
The most interesting situation is realized by aligning the dipolar moments perpendicularly
to the plane where the atoms are confined. The interactions are then purely repulsive
within the plane.
It is likely that
the ``conventional'' quantum Hall states as described by the composite fermion
picture~\cite{Jain89,Lopez91,Kalmeyer92,Heinonen,JainBook} will survive with dipolar interactions. 
However the dipolar interaction induces
interactions between particles with all even relative angular momenta in the LLL. As a 
consequence, the degeneracy when the filling factor is less than 1/2 is lifted
and new physics takes place. Experience with electron systems has shown that one
expect the formation of composite particles in which there is binding of vortices
to the bosonic atoms. It is known~\cite{Regnault03,Chang05,Regnault06} that 
there is formation of such composite objects 
with one vortex per atom leading to the appearance of a series of incompressible liquids
for filling factors $\nu = p/(p+1)$ with $p$ an integer. Similarly for smaller filling factors
one may have formation of composite fermions (CF) made of three vortices plus one boson
giving rise to fractions $\nu =p/(3p\pm 1)$ ($p$ integer). Indeed from some numerical studies
there is limited evidence for such phases although they have not been investigated in detail.

A fascinating property of the quantum Hall phases for bosons is that there are also
states that do not belong to the composite-fermion family. This is the case for complete
filling of the LLL by spinless bosons ($\nu =1$). Here the ground state is described
by the Moore-Read (a.k.a Pfaffian) state which is a paired state of composite fermions. 
This very special state of matter~\cite{Moore91,Greiter92}
supports excitations with non-Abelian statistics and is the object of much attention since
many years. It has been suggested in the context of electronic systems as a plausible
explanation of a quantum Hall state observed for filling factor $\nu =5/2$ of
two-dimensional electron gases. In the case of bosons with purely s-wave scattering,
this state has been observed from exact diagonalizations in the planar geometry
and then confirmed by finding the signature of its paired character in studies on the
spherical geometry. This kind of paired state is in competition with the Fermi sea
made of composite fermions~\cite{Halperin93} and this competition is governed solely by the nature
of interactions. In the case of electrons for example at filling factor
$\nu =1/2$ in the LLL, it is known that a Fermi sea of composite fermions is the ground state
of the system. The second Landau level at electron filling factor $\nu =5/2$
may be approximately regarded as a $\nu =1/2$ electron gas with renormalized 
Coulomb interactions due to the change of electronic wavefunctions sitting on top
on an inert $\nu =2$ background (the filled LLL) provided the cyclotron gap is large enough
compared to interactions. This effective $\nu =1/2$ system has a ground state which
is gapped and may possibly be described by the Pfaffian state.

The case of dipolar atoms offer an additional parameter to tune the interactions. Indeed
there is the s-wave scattering length which can be approximated by a zero-range potential
in addition to the long-range dipolar scattering whose strength is governed by
the dipole moment of the atoms. The ratio of energies corresponding to these two phenomena
can be tuned by Feshbach resonance manipulation which acts upon the s-wave scattering length.
This ratio is a new parameter that can be varied to explore the many-body physics of
ultracold atoms~\cite{Menotti07-1,Menotti07-2}.

In this paper, we explore the competition between the Pfaffian state and the CF Fermi sea
for ultracold atoms in the LLL as a function of the ratio of the dipolar scattering and
the pure s-wave scattering. 
We use exact diagonalizations of the many-body problem for a small number of atoms.
Indeed the restriction to the LLL implies that the Hilbert space of the interacting particles
is finite-dimensional for a finite number of vorticity quanta in the system. Thus
we are led to a (huge) linear algebra problem but without any additional approximation.
This methodology has been very successful in the study of quantum Hall phases.
We consider the spherical as well as torus geometry.

At filling factor $\nu =1$ we find that there is
a wide range of stability of the Pfaffian when s-wave scattering dominates dipolar 
interactions. If we weaken
the latter, there is a phase transition towards a stripe phase with unidimensional density modulation.
Further weakening lead to collapse of the gas phase as in the more conventional three-dimensional
case. A particularly interesting case is $\nu =1/3$. Here one may have formation of composite fermions
with three vortices bound to each atoms instead of one vortex in the $\nu \geq 1/2$ region.
We find that there is a stable Fermi sea without pairing instability i.e. no Pfaffian forms.
This Fermi sea is identified by striking closed shell effects observed by exact
diagonalizations in the spherical geometry. Again, weakening of s-wave scattering leads to collapse
but now without any intervening intermediate phase. The occurrence of this Fermi sea
is a striking phenomenon of emergence of fermions out of bosons in a two-dimensional regime.
This can be exploited to reveal the quantum Hall physics by measuring the occupation number
density. When tuned to this filling factor $\nu=1/3$ it will display the characteristic
signature of a Fermi sea.
Evidence for such compressible state can be obtained by taking a snapshot of the
particle distribution after free expansion as suggested in the work of Read and Cooper~\cite{Read03}.

In section II, we discuss the interactions between atoms in the LLL when they have dipolar interactions
and how this changes the so-called pseudopotentials, the discrete set of energies
that are the solution of the two-body problem in the LLL. In section III,
we present the two geometries we have used for exact diagonalizations~: namely, the sphere and the torus.
Section IV is devoted to the Pfaffian state realized at $\nu =1$, we discuss its stability
and the appearance of the stripe phase.
In section V, we show the appearance of the spectral signature of CFs with one vortex ($^1$CFs) as well
as CFs with three vortices ($^3$CFs) when the filling factor is now 1/3. The formation of $^3$CFs
is shown by looking at the low-lying levels of the many-body problem  in the spherical geometry.
The energy of these low-lying states when fitted against a simple free fermion model
allow to get a estimate of their effective mass which is solely governed by interactions
as it should be in the LLL. Finally section VI contains our conclusions.

\section{Dipolar bosons in the LLL}

We consider a gas of bosonic atoms of mass $M$ confined in a harmonic trap
with cylindrical symmetry along the $z$ axis. The trap frequencies along
the axial and transverse directions are denoted $\omega_{\parallel}$ and
$\omega_{\perp}$ respectively. There are thus associated characteristic
lengths $\ell_{\parallel,\perp}=\sqrt{\hbar/(M\omega_{\parallel,\perp})}$.
We consider the case of a single hyperfine species of bosonic atoms.
At long distances the interaction potential includes a Van der Waals interaction
tail. For ultracold atoms it may be replaced by a simple isotropic
short-range delta-function potential proportional to the s-wave scattering length $a_s$~:
\begin{equation}
 V_s(\mathbf{r})= \frac{4\pi\hbar^2 a_s}{M}\delta^3(\mathbf{r}).
\label{swave}
\end{equation} 
Atoms with a permanent dipole moment also have a dipole-dipole interaction
in addition to the interaction through $V_s$. The corresponding potential is~:
\begin{equation}
\label{dip1}
 V_{dd}(\mathbf{r})= C_{dd}\frac{1-3\cos^2\theta}{r^3}.
\end{equation} 
Here $C_{dd}=d^2/4\pi \epsilon_0$ with $d$ is the electric dipole moment
or  $C_{dd}=\mu_0\mu^2/4\pi$ with $\mu$ the magnetic moment. The angle $\theta$
is between the dipole orientation and the vector joining the two atoms.
We can make a dimensionless parameter out of these two kind of interactions~:
\begin{equation}
 \epsilon_{dd}\equiv\frac{MC_{dd}}{3\hbar^2 a_s}.
\label{epsdip} 
\end{equation} 
In the case of the well-studied $^{87}$Rb, dipolar interactions are negligible
with $\epsilon_{dd}\approx 0.007$. Experiments have achieved Bose-Einstein
condensation of Chromium atoms which have a magnetic dipole moment of $6\mu_B$
and a s-wave scattering length of order one hundred Bohr radius, leading
to a sizeable $\epsilon_{dd}\approx 0.16$. Several Feshbach resonances are known
for some hyperfine states of Chromium that may allow to tune $a_s$ to a small value
of the order of the Bohr radius leading to a big increase of the parameter $\epsilon_{dd}$.
It has been possible to enhance the effect of dipolar interactions up to 
$\epsilon_{dd}\approx 0.8$~\cite{Lahaye07}.

In this paper we study the special case of dipolar atoms in a rotating two-dimensional 
(2D) regime
for which dipolar moments are perpendicular to the plane 
(hence $\theta =\pi /2$ in Eq.(\ref{dip1})). The motion along the $z$-axis
is considered to be frozen i.e. the $z$-dependence of the wavefunctions is
the Gaussian ground state of the harmonic confining potential.
The contact potential Eq.(\ref{swave}) becomes effectively two-dimensional~:
\begin{equation}
 V_s^{2D}(\mathbf{r})=\sqrt{8\pi}\,\, \frac{\hbar^2 }{M}
\frac{a_s}{\ell_{\parallel}}
\delta^2(\mathbf{r}).
\label{Vs} 
\end{equation} 
The dipolar interaction in this 2D regime is given by the following integral~:
\begin{equation}
 V_{dd}^{2D}(x_1-x_2,y_1-y_2)=\int dz_1 dz_2\,\, 
V_{dd}(x_1-x_2,y_1-y_2,z_1-z_2)
\,\, \dfrac{1}{\pi \ell^2_\parallel}\,
{\mathrm e}^{-(z_1^2+z_2^2)/\ell^2_\parallel}.
\label{Vdd} 
\end{equation} 

We concentrate onto the regime where all atoms stay in the LLL. For rotating systems
this may happen by tuning the rotation frequency of trap to the frequency of
the harmonic confining force. The neutral atoms just feel the Coriolis force
which is formally equivalent to a Lorentz force. This correspondence will be
used throughout this paper. We will use both languages (rotating and magnetic)
to clarify the physics of the rotating system. The LLL is spanned by one-body
states of the following form~:
\begin{equation}
 \phi_{m}(z) = \dfrac{1}{\sqrt{2^{m+1}\pi m!}} z^m {\mathrm e}^{-|z|^2/4\ell^2},
\end{equation} 
where $z=x+iy$, the length scale is set by the magnetic length 
$\ell =\sqrt{\hbar/eB}=\sqrt{\hbar/2M\omega_\perp}$ and $m$ is a nonnegative integer
which is (in units of $\hbar$) the angular momentum along $z$ of the state.
The two-body problem can be solved straightforwardly in the LLL when one has
an interacting potential which is rotationally invariant. The eigenenergies
are given by~:
\begin{equation}
 V_m =\int_0^\infty qdq \,\,\tilde{V}(q)\, L_m(q^2\ell^2)\, {\mathrm e}^{-q^2\ell^2},
\end{equation} 
where $\tilde{V}(q)$ is the Fourier transform of the 2D potential and $L_m$ the Laguerre
polynomials. These energies are named pseudopotentials and completely parametrize
the N-body problem. Indeed the interaction Hamiltonian in the LLL can be written as~:
\begin{equation}
 \mathcal{H}=\sum_{i<j}\sum_{m}\,V_m \,\, \mathcal{P}^{(m)}_{ij},
\end{equation} 
where $\mathcal{P}^{(m)}_{ij}$ projects the state of the pair $(ij)$
onto relative angular momentum $m$.
The local potential $V_{s}^{2D}$ Eq.(\ref{Vs}) only contributes to $V_0$ since
it scatters only particles with zero relative angular momentum.

On the contrary, the dipolar potential $V^{2D}_{dd}$ Eq.(\ref{Vdd}) 
contributes to all
$V_m$'s. In the limit of strong confinement along the $z$ axis 
$\ell_\parallel\rightarrow 0$, the dipolar contribution to the $V_m$'s
is finite~\cite{Cooper05} and given by~:
\begin{equation}
 V_m^{dd} = \frac{C_{dd}}{\ell^3}\frac{\sqrt{\pi}}{8}\frac{(2m-3)!!}{2^{m-1}m!}.
\label{dipVm} 
\end{equation} 
On the contrary, the contribution of the dipolar potential
to $V_0$ is diverging in this limit. In fact $V_0$ is a local interaction
which has contributions from both interactions. We consider the
relative ratio to be an adjustable parameter as was realized in the three-dimensional situation
by manipulation of Feshbach resonance. We thus fix all pseudopotentials
for $m\neq 0$ to their value given by Eq.(\ref{dipVm}) and consider the ratio
$\alpha =V_0/V_2$ as a free parameter. Large values of $\alpha$ will give
back the FQHE physics of bosons with contact interactions
and novelties emerge at finite $\alpha$.
The ratio $\alpha$ is a function not only of $C_{dd}$ and $a_s$ but
is also affected by the lengths $\ell$ and $\ell_\parallel$. Its control is then
more complex than the case of bulk nonrotating gases where the competition
is governed by $\epsilon_{dd}$ Eq.(\ref{epsdip}).

\section{Geometries for exact diagonalizations}

To investigate the ground state properties of interacting dipolar bosons
in the LLL, we resort to exact diagonalization techniques. It is convenient
to use geometries without boundaries to access the bulk physics. In this paper
we use both the sphere and the torus geometry. We first briefly discuss
the use of the spherical geometry~\cite{Haldane83,Fano86}. 
The Landau problem on the sphere
leads to a finite number of one-body states in the LLL proportional
to the area of the sphere. 
The spherical geometry possesses the full rotation symmetry and thus
all states can be classified according to their angular momentum
referred to as spin in what follows.
More precisely if an integer number of flux quanta
(vorticity quanta in the rotating language) is piercing the sphere 
$N_\phi \equiv 2S$, its radius is $R=\ell\sqrt{S}$ and the LLL is a set
of 2$S$+1 states forming a spin-S multiplet. In spherical coordinates
$(\theta,\varphi)$, the (unnormalized) one-body wavefunctions can be taken as~:
\begin{equation}
 \Phi^{S}_{M}=u^{S+M}v^{S-M},
\quad u=\cos(\theta/2)\, {\mathrm e}^{-i\varphi/2}
\quad v=\sin(\theta/2) \, {\mathrm e}^{i\varphi/2},
\end{equation} 
where $M=-S,\dots,+S$. The Laughlin wavefunction~\cite{Laughlin83} in the unbounded plane geometry
is given by~:
\begin{equation}
\Psi^{(m)}=\prod_{i<j}(z_i - z_j)^{m}\, {\mathrm e}^{-\sum_i |z_i|^2/4\ell^2}.
\label{laughlinP} 
\end{equation} 
It can be translated on the sphere
by the following formula~:
\begin{equation}
\Psi^{(m)}=\prod_{i<j}(u_i v_j - v_i u_j)^{m}.
\label{laughlinS} 
\end{equation} 
This wavefunction describes a liquid state with filling factor $\nu=1/m$
smoothly covering the sphere (if $m$ is not very large). Even values of $m$
give candidate ground states for bosonic systems. This state is singlet
under rotations and hence non-degenerate on the sphere. It has also a specific
relation between the flux and the number of particles namely
$N_\phi = m(N-1)$ when requiring that all particles are in the LLL.
This geometry is convenient for the study of liquid states
that possess full rotation symmetry but it is not adapted
to the detection of phases with broken translational symmetry
like one or two-dimensional analogs of charge density waves.
In general a state with filling factor $\nu$ will have a 
flux-number relation of the form $N_\phi = N/\nu -X$
where $X$ is a non-trivial shift depending upon the state under consideration.

The other geometry is that of the torus~\cite{Haldane85,Haldane85-2}.
We use a periodic rectangular geometry of sides $a$ and $b$.
Due to the presence of the magnetic field, standard translation
operators do not commute and their implementation is not trivial.
Haldane has shown how to construct eigenstates of the many-body
problem with two conserved pseudomomenta corresponding
to the translations along the two directions. This pseudomomentum
$\mathbf K$ lies in a restricted Brillouin zone which contains
only $N_0^2$ points where $N_0$ is the GCD of the number of particles and
the number of flux quanta. 
For convenience, we measure $K_x$ in units of $2\pi\hbar/a$ and
$K_y$ in units of $2\pi\hbar/b$.
Contrary to the case of the sphere
there is no special shift in the relation between the flux and the number
of particles. There is however a center of mass degeneracy which is
exact for all system sizes and which is given by $q$ at 
filling fraction  $p/q$. We always factor out
this degeneracy in our numerical studies. The efficient coding
of the magnetic translation leads to diagonalization in spaces of reduced
dimension in addition to the possibility to classify states
by their pseudomomentum. In the rectangular case there are also
discrete symmetries relating  states at $(\pm K_x, \pm K_y)$.
The Laughlin wavefunction can be again translated in the torus geometry~\cite{Haldane85-2}
with the help of theta functions~:
\begin{equation}
 \Psi^{(m)}=F_{CM}^{(m)}\,\, (\sum_i z_i)\prod_{i<j}\theta_1(z_i - z_j|\tau )^{m}\, 
{\mathrm e}^{-\sum_i y_i^2/2\ell^2},
\label{laughlinT} 
\end{equation} 
where $\tau =ib/a$ and the center of mass dependence $F_{CM}^{(m)}$ is itself a theta function 
with characteristics.

\section{Pfaffian vs stripe at $\nu=1$}

Many aspects of the physics of the FQHE states can be understood by use of the CF picture.
In this approach one is reasoning with effective CF fermions that are
made of the original particle bound with some number (n) of flux quanta
depending on circumstances. If the bare particle is a boson then we need
an odd number $n$ of flux tubes to convert it into a CF.  The effective
number of flux quanta for $^n$CFs is then $N_\phi^* = N_\phi -nN$.
Since this flux is reduced with respect to the field acting on the bare particles
the $^n$CFs may occupy $p$ higher-lying LLs and we expect commensurability
effects for each integer filling of the effective LLs.
Formation of $^1$CFs leads to a satisfactory explanation of a series
of incompressible states observed numerically at filling factor 
$\nu =p/(p+1)$ for at least $p=1,2,3$. This series of fractions converges
towards $\nu =1$. At this filling factor the $^1$CFs feel zero flux
and one possible candidate for the ground state is a Fermi sea of $^1$CFs
as is realized in electronic systems at $\nu =1/2$. However formation
of a Fermi sea may be preempted by instability towards pairing of CFs.
The possibility of pairing has been proposed in the context of electronic systems
and boils down to the consideration of the so-called Pfaffian wavefunction~\cite{Moore91}~:
\begin{equation}
\Psi_{Pf}={\rm Pf}\left(\frac{1}{z_i-z_j}\right)\prod_{i<j}\left(z_i-z_j\right)
 {\mathrm e}^{-\sum_i |z_i|^2/4\ell^2},
\label{pfaff}
\end{equation} 
where ${\rm Pf}$ stands for the Pfaffian symbol. This latter object is defined for an arbitrary
skew-symmetric $N\times N$ (N even) matrix $A$~:
\begin{equation}
{\rm Pf}\left(A\right)=\sum_{\sigma} \epsilon_{\sigma} A_{\sigma(1)\sigma(2)}
A_{\sigma(3)\sigma(4)}...A_{\sigma(N-1)\sigma(N)},
\label{pfdef}
\end{equation}
where the sum
runs over all permutations of the index with N values and
$\epsilon_{\sigma}$ is the signature of the permutation.
This state can be formulated in the spherical geometry by use
of the substitution $z_i - z_j \rightarrow u_i v_j-u_j v_i$.
This leads to a singlet state which has the special flux-number of particle relationship 
$N_\phi =N-2$. Exact diagonalizations on the disk geometry
have given evidence for this state in the case of pure contact interactions
between bosons~\cite{Wilkin00}. This has been clarified in the spherical
geometry~\cite{Regnault03}.

\subsection{Evidence for Pfaffian state on the torus}

The Pfaffian state at $\nu=1$ has also been observed in the torus geometry~\cite{Cooper01}
for the contact interaction Eq.(\ref{Vs}). Its striking signature is then
the ground state degeneracy which is of topological origin.
This can be seen in several ways. We first note that in the case of the Laughlin state
the Jastrow factor is rendered periodic by simply doing the substitution
$z_i - z_j\rightarrow\theta_1(z_i - z_j|\tau )$ (forgetting about the center of mass motion).
While this is correct for the Jastrow factor in the Pfaffian wavefunction Eq.(\ref{pfaff}),
it cannot be correct for the Pfaffian factor that involves \textit{inverse} powers of $z_i-z_j$.
The correct substitution is in fact~\cite{Greiter92} 
$1/(z_i-z_j)\rightarrow \theta_a(z_i-z_j|\tau)/\theta_1 (z_i-z_j|\tau)$ with $a=2,3,4$.
This leads to \textit{three} ground states~: this degeneracy has nothing to do
with that of the center of mass which is anyway absent for $\nu=1$.
Such a topological degeneracy is expected to be approximate for any finite size system
and should become more and more manifest as we converge towards the thermodynamic limit.

\begin{figure}
\begin{center}
\resizebox{100mm}{!}{\includegraphics{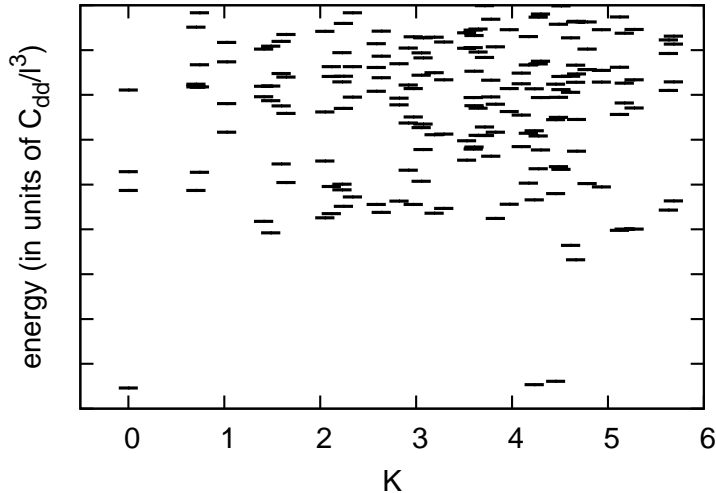}}
\caption{Low-lying energy levels of 12 bosons at $\nu =1$ with dipolar interactions
characterized by $\alpha = V_0/V_2 =3$ as a function of the pseudomomentum $K$. 
The rectangular cell has aspect ratio $a/b=0.95$}
\label{threefold}
\end{center}
\end{figure}

For large values of $\alpha = V_0/V_2$ we find that the Pfaffian phase survives
the addition of dipolar interactions. This is of course what one can expect for a gapped phase
that exists in the $\alpha\rightarrow\infty$ limit. A typical spectrum in this regime
is given in fig.(\ref{threefold}) for N=12 bosons.
We have used a small departure from square 
unit cell to show more clearly
the degenerate states.

The three ground states have ${\mathbf K}=(0,0)$, ${\mathbf K}=(6,0)$ and
 ${\mathbf K}=(0,6)$. These quantum numbers are exactly those expected from 
the bosonic Pfaffian formulated on the torus. These states are separated by a clear
gap from the bulk of the spectrum and this situation is robust with respect
to the change of the aspect ratio. 

\subsection{Evidence for stripe phase}

\begin{figure}
\begin{center}
\resizebox{100mm}{!}{\includegraphics{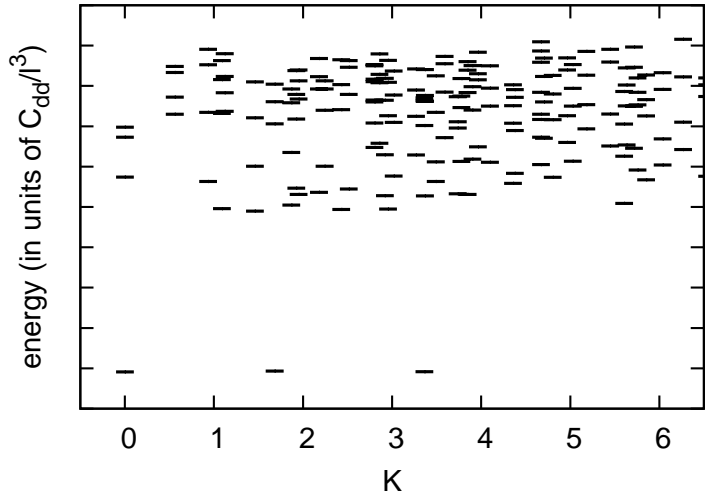}}
\caption{Low-lying energy levels of 12 bosons at $\nu =1$ with dipolar interactions
characterized by $\alpha = V_0/V_2 =2$ as a function of the pseudomomentum $K$. 
The rectangular cell has aspect ratio $a/b=0.6$}
\label{stripe}
\end{center}
\end{figure}

This range of stability of the Pfaffian terminates around
$\alpha \approx 2.4$. Below this critical value we find that the ground state
degeneracy changes and that the whole spectrum looks compressible
contrary to the Pfaffian case. 
In a square unit cell we find ground states at ${\mathbf K}=(0,0),(0,\pm 4),(\pm 4,0)$.
This regime is very sensitive to the aspect ratio of the unit cell contrary to the
Pfaffian phase.
 The degeneracy is lifted in an anisotropic cell towards
${\mathbf K}=(0,0),(0,\pm 4)$. These wavevectors are now inside the magnetic Brillouin
while the Pfaffian ground states were at half reciprocal lattice vectors.
This set of vector form a 1D array, they are separated by a characteristic 
ordering wavevector ${\mathbf Q}=(0,3)$. This special configuration
persists for a wide range of aspect ratios $0.4\lesssim a/b \lesssim 0.9$.
This is exactly what we expect for the formation
of an unidimensional density modulation or stripe phase~\cite{Rezayi99,Haldane00,Yang01}.
A typical spectrum in this phase is shown in Fig.(\ref{stripe})
for N=12 bosons.

If we now further reduce the parameter $\alpha$ there is a collapse
of the system of bosons for $\alpha \approx 1$ below which there is
a huge structureless degeneracy of the ground state. Although we have
not studied in detail this non-FQHE phase, we believe it is the LLL
analogous of the collapse of three-dimensional non-rotating gases
observed when the parameter $\epsilon_{dd}$ becomes less than 
one~\cite{Menotti07,Menotti07-1,Menotti07-2}. It is worth stressing that
we have no evidence for other intermediate phases between the Pfaffian
and the collapsed phase. This is contrary to what is known~\cite{Rezayi05} in the
case of the Laughlin state at $\nu =1/2$. Here the Laughlin state
has also a wide range of stability. For small values of $\alpha$
it is replaced by a striped phases as we find for $\nu =1$
but further reduction leads to so-called bubble phases
i.e. regular 2D arrays of density excess with two or more
bosons per site.

\section{$^3$CF Fermi sea at $\nu=1/3$}

We now turn to the fate of the bosons interacting with dipolar
interactions when the filling factor is tuned to 1/3. This is
also a very special case. The composite fermion construction
certainly suggests that at small Bose filling factors
there may be formation of $^3$CFs. This leads to the possibility
of standard incompressible liquids for $\nu=p/(3p\pm 1)$ that would
be the Bose analogs of the series of $^4$CFs states observed
for electrons for $\nu < 1/3$. There is in fact some evidence reported
in the literature for such fractions~\cite{Cooper05,Regnault03}.
This series has also a termination point at $\nu =/1/3$ at which the
$^3$CFs feel zero effective flux and we are again facing the competition
between the Fermi sea of CFs and the Pfaffian state.
To construct a Pfaffian candidate~\cite{Moore91} at 1/3 it is enough to add two
powers to the Jastrow factor in Eq.(\ref{pfaff}) leading to~:
\begin{equation}
\Psi_{Pf}={\rm Pf}\left(\frac{1}{z_i-z_j}\right)\prod_{i<j}\left(z_i-z_j\right)^3
 {\mathrm e}^{-\sum_i |z_i|^2/4\ell^2}.
\label{pfaff3}
\end{equation} 
Translated on the sphere it leads to a series of incompressible states with
$N_\phi = 3N-4$. ON the torus it has also a topological degeneracy. This
is not what we observe. In fact we find no evidence for the Pfaffian
neither on the sphere nor on the torus for any values of $\alpha$.
Our findings are compatible with the presence of a Fermi sea of $^3$CFs.
The wavefunction of such a state on the disk geometry is~\cite{Rezayi94}~:
\begin{equation}
 \mathcal{P}_{LLL}\{\prod_{i<j}\left(z_i-z_j\right)^3 {\mathrm {det}}[
\exp (i{\mathbf k}_i \cdot {\mathbf r}_j)]\}.
\end{equation} 
In this equation the Jastrow factor ensuring filling factor 1/3 is multiplied
by the Slater determinant of plane waves which is describing the Fermi sea.
Since this determinant is not automatically in the LLL one needs to project it onto
the LLL explicitly. The choice of the $\mathbf k$ vectors in the determinant
to find the ground state is dependent upon details of the geometry
of the system. The signature of such a state is that the ground state
pseudomomentum in the torus geometry is not as simple as in the case
of FQHE states or striped phases we have encountered above.
It is thus much simpler to study the properties of the Fermi sea
on the spherical geometry.

\subsection{$^1$CFs versus $^3$CFs}

\begin{figure}
\begin{center}
\resizebox{100mm}{!}{\includegraphics{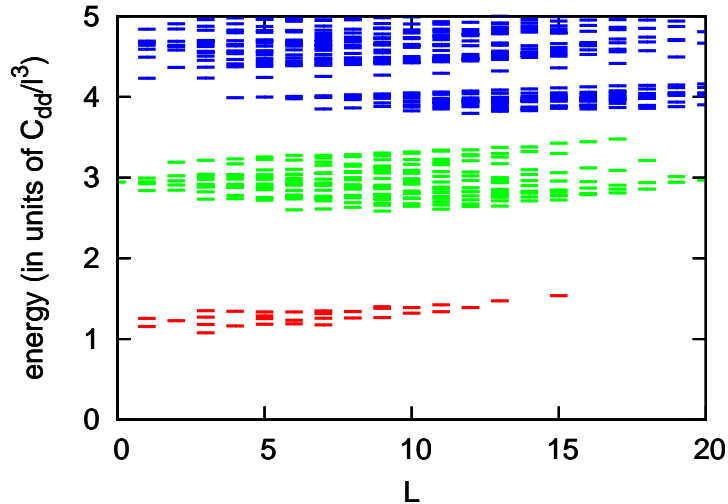}}
\caption{(Color online) The lower part of the spectrum of 6 bosons on the sphere at 
$N_\phi=15$ with dipolar interactions
characterized by $\alpha = V_0/V_2 =10$, $V_0=2.73$ and $V_2=0.273$ as a function of the total
angular momentum $L$. The lowest band of states at energies close
to one unit avoids relative angular momentum zero, the next band of states
has on average one pair with zero momentum anf hence the energy scale is raised
by $\approx V_0$, the counting extends to higher levels.
The low-energy manifold is in one-to-one correspondence with the spectrum
of 6 \textit{electrons} at $N_\phi = 10$ i.e. at $\nu =1/2$.}
\label{cfs}
\end{center}
\end{figure}

We first discuss the nature of the correlations between bosons at filling factors
smaller than 1/2. The first point is that the ground state becomes highly
degenerate when we are dealing with the pure contact interaction Eq.(\ref{Vs}).
Indeed if we multiply the Laughlin wavefunction Eq.(\ref{laughlinP}) for
$\nu =1/2$ by any symmetric polynomial of the $z_i$ coordinates it 
is still trivially a zero-energy eigenstate. Some of these states are edge
modes when the degree of the polynomial is small enough but this also extends
to states with smaller filling factors i.e. the $\nu=1/4$ Laughlin state is itself
lost in the manifold of such states. These states do not feel the delta
interactions since the relative angular momentum between any two bosons
is at least two. However excited states, hence with nonzero energies,
feel the effect of the pseudopotential $V_0$. We thus expect to find
above the ground state a set of states corresponding to many-body states
in which there is one pair of bosons having relative angular
momentum zero~: such states will have energy of order $V_0$. There will then
yet another band with \textit{two} pairs of bosons interacting with $V_0$
and energy of order $2V_0$ and so on. These quasi-degeneracies are most clearly exhibited
in the spherical geometry for which they are not further complicated
by the center of mass degeneracy that exists on the disk.
If we now introduce some dipolar interactions all the pseudopotentials
$V_m, m\geq 2$ become nonzero. The ground state manifold is now no
longer degenerate and will split into subbands according
to the number of pairs of bosons having relative angular momentum 2 at least
and so on. This picture of classification by relative angular momentum
has been put forward some time ago by Wojs and Quinn~\cite{Wojs00,Quinn00}
for electrons interacting by the Coulomb potential. In their case the separation
of scales is not tunable while the dipolar system offers an opportunity
to observe the build-up of these hierarchical structures. 
Since the pair angular momentum is not a conserved quantity one should keep in mind
that the
classification is approximate. The quantum numbers of these various sets
of states can be found by using 
the language of composite fermions~\cite{Peterson04}.
The $^1$CFs feel an effective flux $N_\phi^* = N_\phi -(N-1)$.
At filling factor 1/3 the configuration of these fermions is thus highly
degenerate. Let us take the example of N=6 bosons given in Fig.(\ref{cfs}).
The $^1$CF LLL has now 11 states. The values of the total angular
momentum for 6 fermions in an 11-fold degenerate shell are exactly
those found in the lowest energy band of Fig.(\ref{cfs}). If we promote
one $^1$CF to the N=1 $^1$CF LL then the momentum of the states we generate
in this configuration are those of N=5 fermions in a 11-fold degenerate shell
and one extra fermion in a 13-fold shell. This is the exact content
of the second band of states in Fig.(\ref{cfs}). The energy scale of
the $^1$CF LLs is related to $V_0$ and it is only the remaining pseudopotentials
$V_m,m\geq 2$ that will lift the degeneracy.
If we look at the band of states predicted by the $^1$CF reasoning
we find indeed that the degeneracy is lifted
by the dipolar force. Now the ordering of energy levels
in the ground state manifold of $^1$CF is predicted
by introducing $^3$CF that predict correctly the smaller splitting
due to the dipolar interaction. 
There is formation of a set of effective $^3$CFs LL with an energy
separation governed essentially by the pseudopotential beyond
$V_0$. We render this reasoning quantitative in the next section by evaluating
the effective mass of the $^3$CF as a function of $\alpha$.

In this peculiar physical system,
the ratio of the energy scales corresponding to $^1$CF and $^3$CF
correlation is tunable. This is a very interesting situation
that is able to shed light on the coexistence of the two species of composite
particles
$^1$CF and $^3$CF since we can in principle separate them.
This means that one can tune the effective CF Landau levels energy scales
separately for the $^1$CFs and for the $^3$CFs.
Similar coexistence effects have been experimentally observed~\cite{Hirjibehedin03}
in the context of two-dimensional electron gases at filling factor
smaller than 1/3. In this context, these experiments have given evidence
for coexistence of $^2$CFs and $^4$CFs excitations. Since the electrons
are interacting through the Coulomb interaction, there is not a huge
difference between the pseudopotentials $V_1$ and $V_3$ corresponding
to these two kinds of correlations. The dipolar system renders the corresponding
separation manifest.

\subsection{Shell effects on the sphere}

\begin{figure}
  \begin{center}
    \begin{tabular}{cc}
      \resizebox{65mm}{!}{\includegraphics{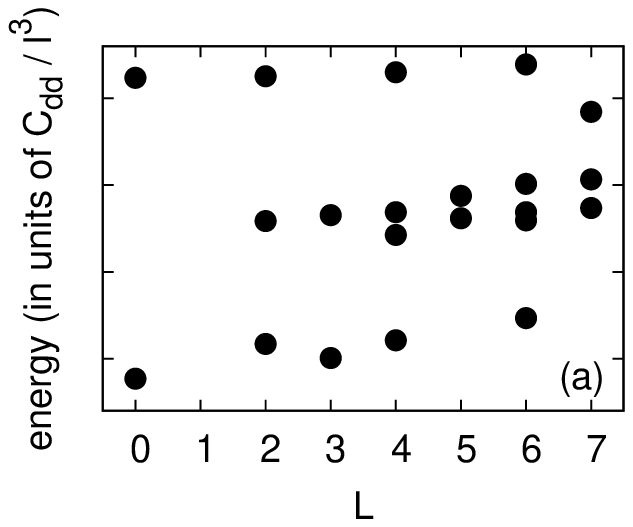}} &
      \resizebox{65mm}{!}{\includegraphics{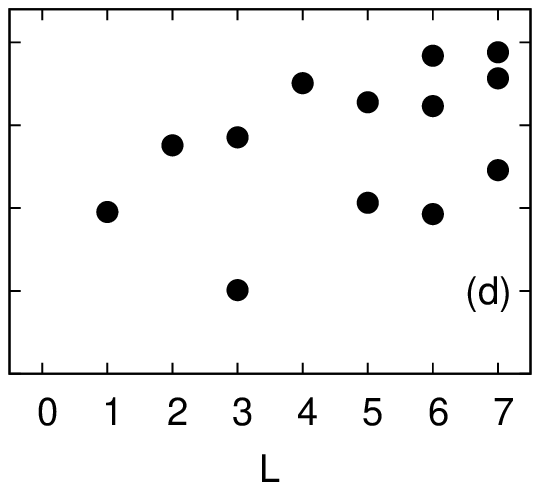}} \\
      \resizebox{65mm}{!}{\includegraphics{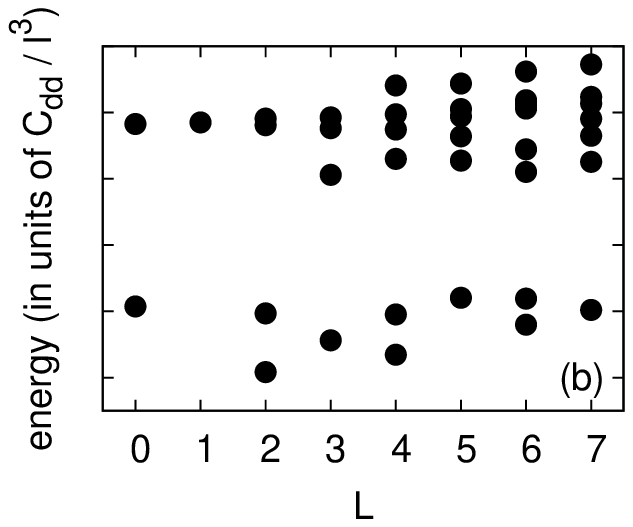}} &
      \resizebox{65mm}{!}{\includegraphics{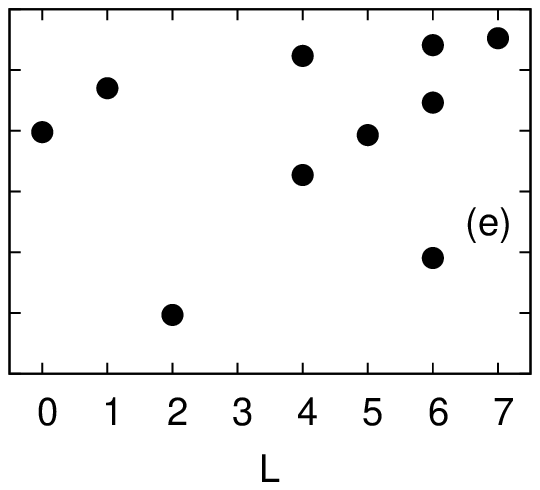}} \\
      \resizebox{65mm}{!}{\includegraphics{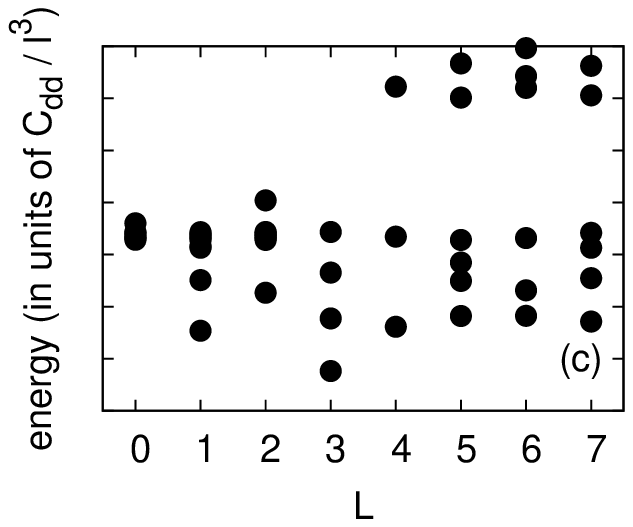}} &
      \resizebox{65mm}{!}{\includegraphics{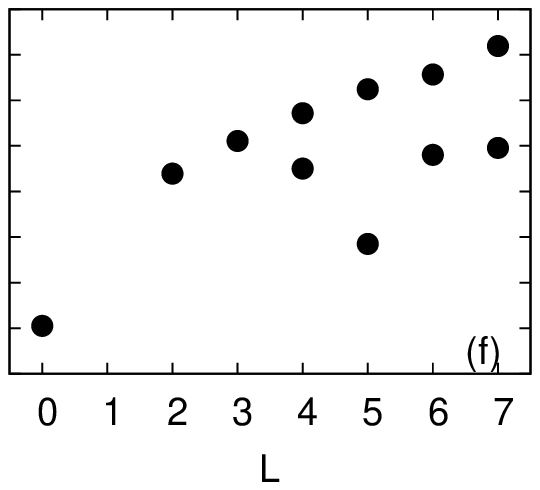}} \\
    \end{tabular}
    \caption{Shell effects of $^3$CFs on the sphere for $\alpha =4$.
     We follow the zero-flux line $N_\phi = 3(N-1)$.
    The low-lying states are plotted as a function of total angular momentum.
     (a) N=4, closed $s$+$p$ shells of $^3$CFs,
     (b) N=5~: one $^3$CF in a $d$ shell leads to $L=3$ ground state,
     (c) N=6~: two $^3$CFs couple to $L=3$,
     (d) N=7~: two holes in the $d$ shell of $^3$CFs lead also to $L=3$,
     (e) N=8~: one hole is similar to the N=5 case,
     (f) N=9~: The three closed shell singlet state is recovered.}
    \label{Shell}
  \end{center}
\end{figure}

The ultimate low-energy behavior of the Bose system is governed by the
lowest-lying states described by the $^3$CFs if there is formation of a Fermi sea.
We have studied the zero-flux line in the $N_\phi - N$ plane, namely
$N_\phi =3(N-1)$. Since the $^3$CF are more or less freely moving on a sphere
we expect to observe closed shell effects when $N$ is a square as it should be
if the effective kinetic energy of the CFs is of the form $L^2$ with $L$ the total
momentum. This is what happens for electrons~\cite{Rezayi94,Morf95,Onoda00,Onoda01}
 in the LLL at $N_\phi =2(N-1)$. When $N=p^2$, $p$ integer, on the zero-flux line the system
ca be considered as a member of the two families of incompressible states with
$\nu=p/(3p\pm 1)$. Indeed the flux-number of particles for these two Jain-like series of states
are given by~:
\begin{equation}
 N_\phi =\frac{3p\pm 1}{p}N \mp p -3.
\label{fluxes} 
\end{equation} 
Along the line belonging to the fraction $p/(3p+1)$ the effective flux
is negative~\cite{Moller05} for $N<p^2$, vanishes at $N=p^2$ and becomes positive beyond
that value. Between two successive squares, Rezayi and Read~\cite{Rezayi94}
have shown that the signature of a CF Fermi sea state is that the angular
momentum of the ground state should be given by the second Hund's rule
for electrons in zero field. This is exactly what we observe between
N=4 and N=9 bosons in Figs.(\ref{Shell},(a)-(f)). At N=4 we have filled $s$
and $p$ shells hence a singlet ground state. At N=5 there is one CF
in the $d$ shell, hence $L=2$. At N=6, the two CFs in the $d$ shell
couple to $L=3$. The sequence is then reproduced backwards by CF particle-hole
symmetry up to the next closed shell configuration~: N=7 has also $L=3$,
N=8 has one hole in $d$-shell hence $L=2$ and N=9 has $s,p,d$ filled shells.
The ordering is similar to that of electrons in the LLL at $N_\phi =2(N-1)$.
At a qualitative level it is exactly what is expected from the formation of $^3$CFs.
One can also recognize the excitations involving effective CF LLs.
For example at N=4, the promotion of one CF from the filled $p$ shell
to the empty $d$ shell leads to the branch of states with $L=2,3,4$.
Similarly at N=9, promotion of one CF from the $d$ shell to the next
$f$ shell leads to a branch of states terminating at $L=5$.

While this is readily analyzed, we find that the torus geometry
leads to a complex dependence of the ground state wavevector as a function
of the aspect ratio and size. This is coherent with a appearance of a 
compressible state. In addition we don't find evidence for any ordering
wavevector like at $\nu=1$ so we rule out the possibility of a striped phase.
Due to the exponential growth of the Hilbert, we are not able to confirm
the level pattern up to the next closed shell ($f$) which should appear 
for N=16 bosons (prohibitively large for exact diagonalizations).

\subsection{Effective mass}

One can now give a more quantitative treatment of the ``free'' $^3$CF picture
by evaluating the effective mass entirely due to the interactions.
The idea is to reproduce ground state energies between closed shell configurations
by a simple free fermion model where the particles are living on the surface
of the sphere. Energies are thus given by~:
\begin{equation}
 E(\{l_i\})= \frac{\hbar^2}{2m^* R^2} \sum_{i=1}^{N}l_i (l_i+1),
\label{freeCF} 
\end{equation} 
where $R=\ell \sqrt{S}$ is the radius of the sphere, the $l_i$
are the momenta of one-body occupied states and the effective mass
is proportional to the interactions for dimensional reasons~:
\begin{equation}
\frac{\hbar^2}{m^*}=\mu (\alpha) \frac{C_{dd}}{a},
\label{mass} 
\end{equation} 
where $a$ is the ion-disk radius, conveniently defined from
electronic systems by $\pi a^2 =1/$density, and $\mu$ is dimensionless. 
As usual in the spherical geometry
there is a difference between the actual density of particles
and the density at the thermodynamic limit for a given filling factor.
This is due to the nonzero shift in the flux-number of particle relation.
While negligible for large numbers, it is sizeable for the values of N at hand.
We correct for this effect by renormalizing the magnetic length or ion-disk radius
by a factor $\sqrt{\nu N_\phi /N}$.

\begin{figure}
\begin{center}
\resizebox{100mm}{!}{\includegraphics{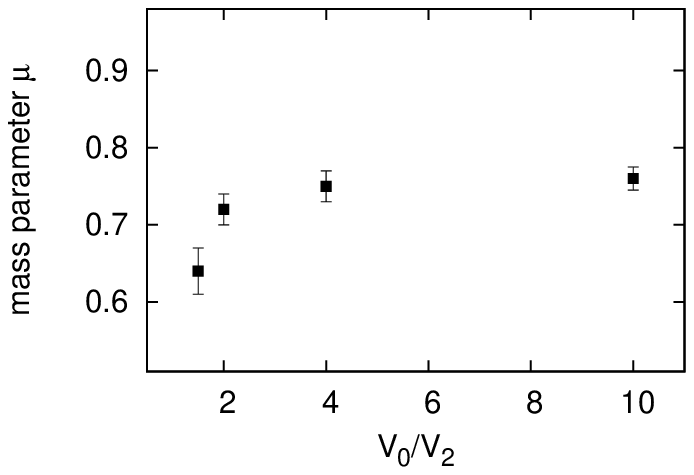}}
\caption{The inverse effective mass parameter $\mu$ of the $^3$CF at $\nu =1/3$
obtained from the ground state energies for $4\leq N \leq 9$ bosons.
The error bars are tentative estimates of the systematic error due to
the fact that real spectra deviate from the free fermion predictions.
As expected from three-flux composite fermions, the effective mass
saturates  for large $\alpha$ to a value $\approx 0.75$ which is independent of the hard-core
pseudopotential $V_0$.}
\label{massfig}
\end{center}
\end{figure}

We find that the energies can be reproduced by Eq.(\ref{freeCF}) as soon as
$\alpha\gtrsim 1$. For lower values the system is in the collapsed phase.
When increasing the hard-core interaction through $V_0$ we find that the
effective mass becomes constant with $\mu \approx 0.75$~: see Fig.(\ref{massfig}). This is exactly
what we expect from the separation of scales between $^1$CF and $^3$CFs.
The degeneracy of the $^3$CF manifold of states at $\nu =/1/3$ is lifted
by the interactions beyond the pure hard-core.

We now comment on the possible strategy to characterize such compressible states.
Indeed the incompressible liquid states 
of the FQHE picture are featureless, i.e. no snapshot of the density
will reveal their character. One has thus resort to more complex
techniques~\cite{Read03,Altman04} to measure correlation functions as a function of the distance
to possible indicators. Also the spatial density variation should display
a step-like structure~\cite{Cooper04} with pinning at magic densities.
The Fermi sea that we have found is a gapless, compressible state.
Since we are in the LLL the signel snapshot measurement of the density
after free expansion reveals directly the particle distribution
as pointed out by Read and Cooper~\cite{Read03}.
One may then observe the density oscillations in space as expected for
a Fermi liquid-like state.

\section{Conclusions}

We have investigated the behavior of bosons in the LLL with dipolar interactions.
The special filling factors we choose on general grounds are expected
to display quantum states different from the Laughlin-like
incompressible states found for example at $\nu =1/2$.
At $\nu =1$ we find that there is an extended stability region of the
Pfaffian state already known for pure contact interactions.
If we weaken enough the hard-core repulsion, this state is destroyed
and replaced by a unidimensional density modulation, a stripe.
In both cases it is the ground state degeneracy which is the main evidence
for the occurrence of such phases. We have studied the fraction $\nu =/1/3$
which is nontrivial in presence of dipolar interactions. Since this state
is the termination point of the Jain-like series of fractions at 
$p/(3p\pm 1)$, it is plausible that there is formation of a CF Fermi sea
as is observed for electron systems at the electronic filling factor 1/2.
We have given evidence for such a state and evaluated the effective
mass of the CFs that forms and give a low-energy approximate description of
the system. The dipolar system is a very interesting case of adjustable
separation between the energy scales involved in the formation
of $^1$CF versus $^3$CFs. The mass at $\nu =1/3$ is remarkably insensitive
to the hard-core repulsion.

Finally we note that formation of fermions out of bosons
is an interesting phenomenon in the realm of the dipolar bosons.
While this phenomenon is well-documented in the one-dimensional context
through the Tonks-Girardeau regime,
it remains to be observed in two space dimensions.

\begin{acknowledgments}

We acknowledge useful discussions with D. Petrov and G. Shlyapnikov.
This work was supported in part by Institut Francilien de Recherche
sur les Atomes Froids (IFRAF).

\end{acknowledgments}






\begin{thebibliography}{99}

\bibitem{Madison00}
K. W. Madison , F. Chevy, W. Wohlleben, and J. Dalibard,
Phys.Rev. Lett. \textbf{84}, 806 (2000);
F. Chevy, K. W. Madison, and J. Dalibard,
Phys. Rev. Lett. \textbf{85}, 2223 (2000)

\bibitem{Aboshaeer01}
J. R. Abo-Shaeer , C. Raman, J. M. Vogels, and W. Ketterle,
Science \textbf{292}, 476 (2001).


\bibitem{BlochRMP}
I. Bloch, J. Dalibard, W. Zwerger,
\rmp , in press 2007.

\bibitem{Wilkin98}
N. K. Wilkin, J. M. F. Gunn and R. A. Smith,
Phys. Rev. Lett. {\bf 80}, 2265 (1998).

\bibitem{Wilkin00}
N. K. Wilkin and J. M. F. Gunn,
Phys. Rev. Lett. {\bf 84}, 6 (2000).

\bibitem{Cooper01}
N. R. Cooper, N. K. Wilkin, and J. M. F. Gunn,
Phys. Rev. Lett. {\bf 87}, 120405 (2001).

\bibitem{schweikhard04}
V. Schweikhard, I. Coddington, P. Engels, V. P. Mogendorff, and E. A. Cornell,
Phys. Rev. Lett. \textbf{92}, 040404 (2004).


\bibitem{Griesmaier05}
A. Griesmaier, J. Werner, S. Hensler, J. Stuhler, and T. Pfau,
\prl \textbf{94}, 160401 (2005).

\bibitem{Stuhler05}
J. Stuhler, A. Griesmaier, T. Koch, M. Fattori, T. Pfau, S. Giovanazzi, 
P. Pedri, and L. Santos,
\prl \textbf{95}, 150406 (2005).

\bibitem{Lahaye07}
T. Lahaye, T. Koch, B. Fr\"olich, M. Fattori, J. Metz, A. Griesmaier,
S. Giovanazzi, and T. Pfau,
Nature \textbf{448}, 672 (2007).

\bibitem{Menotti07}
C. Menotti, C. Trefzger, and M. Lewenstein,
Phys. Rev. Lett. \textbf{98}, 235301 (2007).

\bibitem{Menotti07-1}
C. Menotti and M. Lewenstein,
``Ultra-cold dipolar gases'',
e-print arXiv/0711.3406.

\bibitem{Menotti07-2}
C. Menotti, M. Lewenstein, T. Lahaye, and T. Pfau,
``Dipolar interaction in ultra-cold atomic gases'',
e-print arXiv/0711.3422.

\bibitem{Read03}
N. Read and N. R. Cooper,
\pra \textbf{68}, 035601 (2003).

\bibitem{Komineas06}
S. Komineas and N. R. Cooper,
\pra \textbf{75}, 023623 (2007).

\bibitem{Cooper05}
N. R. Cooper, E. H. Rezayi, and S. H. Simon,
\prl  \textbf{95}, 200402 (2005) 

\bibitem{Zhang05}
J. Zhang and H. Zhai,
\prl \textbf{95}, 200403 (2005).

\bibitem{Rezayi05}
E.H. Rezayi, N. Read, and N. R. Cooper
\prl \textbf{95}, 160404 (2005) 

\bibitem{Cooper07}
N. R. Cooper and E. H. Rezayi
\pra \textbf{75}, 013627 (2007) 


\bibitem{Jain89}
J.K. Jain,
Phys. Rev. Lett. {\bf 63}, 199 (1989).

\bibitem{Lopez91}
A. Lopez and E. Fradkin,
Phys. Rev. B\textbf{44}, 5246 (1991).

\bibitem{Kalmeyer92}
V. Kalmeyer and S. C. Zhang,
Phys. Rev. B\textbf{46}, 9889 (1992).


\bibitem{Heinonen}
\textit{Composite Fermions : A Unified View of the Quantum Hall Regime},
edited by O. Heinonen (World Scientific, Singapore, 1998).

\bibitem{JainBook}
J. K. Jain,
\textit{Composite Fermions}, Cambridge University Press, Cambridge, 2007.


\bibitem{Regnault03}
N. Regnault and Th. Jolicoeur,
Phys. Rev. Lett. {\bf 91}, 030402 (2003);
Phys. Rev. B{\bf 69}, 235309 (2004).

\bibitem{Chang05}
C. C. Chang, N. Regnault, Th. Jolicoeur and J. K. Jain,
Phys. Rev. A{\bf 72}, 013611 (2005).

\bibitem{Regnault06}
N. Regnault, C.~C. Chang, Th. Jolicoeur, and J.~K. Jain
J. Phys. B~: At. Mol. Opt. Phys. \textbf{39} S89 (2006).

\bibitem{Moore91}
G. Moore and N. Read,
Nucl. Phys. B{\bf 360}, 362 (1991).

\bibitem{Greiter92}
M. Greiter, X. G. Wen, and F. Wilczek,
Nucl. Phys. B\textbf{374}, 567 (1992).

\bibitem{Halperin93}
B. I. Halperin, P. A. Lee, and N. Read,
\prb \textbf{47}, 7312 (1993).

\bibitem{Haldane83}
F. D. M. Haldane,
Phys. Rev. Lett. {\bf 51}, 605 (1983).

\bibitem{Fano86}
G. Fano, F. Ortolani, and E. Colombo,
Phys. Rev. B\textbf{34}, 2670 (1986).

\bibitem{Laughlin83}
R. B. Laughlin,
Phys. Rev. Lett. {\bf 50}, 1395 (1983).

\bibitem{Haldane85}
F. D. M. Haldane,
Phys. Rev. Lett. \textbf{55}, 2095 (1985).

\bibitem{Haldane85-2}
F. D. M. Haldane and E. H. Rezayi,
\prb \textbf{31}, 2529 (1985).

\bibitem{Rezayi99}
E. H. Rezayi, F. D. M. Haldane, and K. Yang,
Phys. Rev. Lett. \textbf{83}, 1219 (1999).

\bibitem{Haldane00}
F. D. M. Haldane, E. H. Rezayi, and K. Yang,
Phys. Rev. Lett. \textbf{85}, 5396 (2000).

\bibitem{Yang01}
K. Yang, F. D. M. Haldane, and E. H. Rezayi,
Phys. Rev. B\textbf{64}, 081301(R) (2001).

\bibitem{Wojs00}
A. Wojs and J. J. Quinn,
Philos. Mag. B\textbf{80}, 1405 (2000).

\bibitem{Quinn00}
J. J. Quinn and A. Wojs,
J. Phys. Condens. Matter \textbf{12}, 265(R) (2000);
Physica (Amsterdam) \textbf{6E}, 1 (2000).


\bibitem{Peterson04}
M. R. Peterson and J. K. Jain,
Phys. Rev. Lett. \textbf{93}, 046402 (2004).

\bibitem{Hirjibehedin03}
C. F. Hirjibehedin, A. Pinczuk, B. S. Dennis, L. N. Pfeiffer,
and K. W. West,
Phys. Rev. Lett. \textbf{91}, 186802 (2003).

\bibitem{Rezayi94}
E. H. Rezayi and N. Read,
\prl \textbf{72}, 900 (1994).

\bibitem{Morf95}
R. Morf and N. d'Ambrumenil,
Phys. Rev. Lett. \textbf{74}, 5116 (1995).

\bibitem{Onoda00}
M. Onoda, T. Mizusaki, T. Otsuka, and H. Aoki,
\prl \textbf{84}, 3942 (2000).

\bibitem{Onoda01}
M. Onoda, T. Mizusaki, and H. Aoki,
\prb \textbf{64}, 235315 (2001).


\bibitem{Moller05}
G. M\"oller and Steven H. Simon,
Phys. Rev. B\textbf{72}, 045344 (2005).

\bibitem{Altman04}
E. Altman, E. Demler, and M. D. Lukin,
\pra \textbf{70}, 013603 (2004).

\bibitem{Cooper04}
N. R. Cooper, F. J. M. van Lankvelt, J. W. Reijnders, and K. Schoutens,
\pra \textbf{72}, 063622 (2005).



\end{thebibliography}
\end{document}